\documentstyle[11pt,LFnewpasp2,twoside,epsf]{article}
\def\edcomment#1{\iffalse\marginpar{\raggedright\sl#1\/}\else\relax\fi}
\reversemarginpar

\begin{document}
\title{The evolution of AGNs in the Hard X-Rays and the Infrared}
 \author{Fabio La Franca, Israel Matute}
\affil{Dipartimento di Fisica, Universit\`a degli studi ``Roma Tre'',
Via della Vasca Navale 84, I-00146 Roma, Italy}
\author{Fabrizio Fiore}
\affil{Osservatorio Astronomico di Roma, Via Frascati 33, I-00040 Monteporzio, Italy}
\author{Carlotta Gruppioni, Francesca Pozzi, Cristian Vignali}
\affil{Osservatorio Astronomico di Bologna, Via Ranzani 1, I-40127 Bologna, Italy}
\author{on behalf of the HELLAS and ELAIS consortii\footnotemark[1]}
\footnotetext[1]{HELLAS:
A. Antonelli, A. Comastri, P. Giommi, G. Matt, S. Molendi,
G.C. Perola, F. Pompilio. ELAIS: C. Lari,
M. Mignoli, G. Zamorani, D. Alexander, S. Oliver, S. Serjeant, A. Franceschini,
G. Danese, M. Rowan-Robinson.}

\begin{abstract}
We present the estimate of the evolution of type 1 AGNs in the hard
(2-10 keV) X-rays drawn from the HELLAS survey, and in the IR (15
$\mu$m) obtained from the ELAIS survey.  We find that the local
luminosity function (LF) of AGN1 in the 2-10 keV is fairly well
represented by a double-power-law-function.  There is evidence for
significant cosmological evolution according to a pure luminosity
evolution model $L_X(z)$$\propto$$(1+z)^k$, with
$k$=$2.12^{+0.13}_{-0.14}$ and $k$=$2.19^{+0.13}_{-0.14}$ in a
($\Omega_{\mathrm{m}}$,$\Omega_\Lambda$)=(1.0,0.0) and in a
($\Omega_{\mathrm{m}}$,$\Omega_\Lambda$)=(0.3,0.7) cosmology
respectively. In a ($\Omega_{\mathrm{m}}$,$\Omega_\Lambda$)=(1.0,0.0)
Universe the data show an excess of faint high redshift type 1 AGN
which is well modeled by a luminosity dependent density evolution,
similarly to what observed in the soft X-rays.  In the IR band, with a
($\Omega_{\mathrm{m}}$,$\Omega_\Lambda$)=(1.0,0.0) cosmology, the
evolution is similar to what observed at other wavebands, the LF is a
double-power-law-function with a bright slope 2.9 and a faint slope
1.1, following a pure luminosity evolution model with
$k$=$3.00^{+0.16}_{-0.20}$.
\end{abstract}

\section{Introduction}

Before the advent of the last generation of hard X-ray telescopes, AGN
samples were predominantly based on type 1 objects
(AGN1\footnote[2]{We here define AGN1 all the AGNs having a broad
emission line spectrum in the optical.}) selected either in the
optical or later on in the soft X-rays by ROSAT. In these bands the
evolution of AGN1 has been well measured (see e.g. Della Ceca et
al. 1992; Boyle et al. 2000; Miyaji, Hasinger and Schmidt 2000). On
the contrary the generation of samples of type 2 AGNs (AGN2) has been
difficult at any wavelength and limited to few local surveys.

The general picture was in favor of a model in which type 1 objects
were associated to AGNs with low absorption in the X-rays while type 2
to obscured sources with large column densities and spectra strongly
depressed in the soft X-rays and emitting mainly in the a) hard X-rays
and b) mid and far infrared.  For these reasons, in order to achieve a
more exhaustive picture of the evolution of the whole family of AGNs,
the researches have been recently concentrated in these two wavebands.

Thanks to their high angular resolution ($\sim$1-5 $''$) the first
spectroscopic identification projects of faint {\it Chandra} and {\it
XMM-Newton} hard X-ray sources have been able to observe faint
(I$\sim$23) optical counterparts. At variance with the classical
type-1/type-2 model in the optical, a significant number of the
counterparts ($\sim$30\%) resulted to be apparently optical normal
galaxies, and moreover part of the optical type 1 AGNs resulted to be
absorbed in the hard X-rays (see e.g. Fiore et al. 2000; Comastri et
al. 2001; Barger et al. 2001).

These observations have complicated the picture of the AGN model. In
the framework of the computation of the density of AGNs, it is not
clear how to handle the classification of the sources and to take
into account the selection biases introduced by the observation in the
2-10 keV range where the absorption still play a relevant role.

At variance, the situation in the mid infrared is in a less advanced
stage because of the scarcity of satellite missions, the only ones
being {\it IRAS} and {\it ISO}. Although the {\it ISO} mission has
been completed a couple of years ago, the only existing survey of AGN
in the mid infrared is still the sample from Rush, Malkan and Spinoglio
(RMS, 1993) obtained from the IRAS observations of the whole sky at 12
$\mu m$ down to $\sim$300 mJy.  We have thus decided to undertake two
projects.

a) While the recent deep surveys with {\it Chandra} and {\it
XMM-Newton} have reached fluxes $\sim 2\times 10^{-16}
erg~cm^{-2}s^{-1}$ (2-8 keV, Brandt et al. 2001) in quite small areas
(less than 1 deg$^2$), at brighter fluxes ($\sim 10^{-13}
erg~cm^{-2}s^{-1}$; 5-10 keV) the density of sources is low (about
5/deg$^2$) and tenths of square degrees are to be covered in order to
collect statistical significant samples. Such samples are quite rare
at moment but are essential in order to fully cover the X-ray
luminosity/redshift plane and study the AGN evolution. In the first
part of this paper we report the results of the spectroscopic
identifications of one of such samples (HELLAS), which covers from
0.2 to 55 deg$^2$ in the flux range $5\times 10^{-14} -5\times
10^{-11} erg~cm^{-2}s^{-1}$ (5-10 keV).

b) In order to extend at deeper fluxes the coverage of the luminosity
redshift plane in the mid infrared (15 $\mu m$) we have carried out
the European Large Area ISO Survey (ELAIS) over 12 deg$^2$ down to
$\sim$0.5-1 mJy, more than 2 order of magnitudes deeper than the
previous RMS sample. In the second part of this paper we will report on
the results of the spectroscopic identifications of this project.

\section{HELLAS: The evolution of AGN1 in the Hard X-ray}

\subsection{The HELLAS survey}

The X-ray sources have been detected by the {\it BeppoSAX}-MECS
instruments in the 5-10 keV band in the framework of the High Energy
Large Area Survey (HELLAS).  Preliminary results have been presented
in Fiore et al. (1999). The whole survey and the catalogue used in this study is
described by Fiore et al. (2001). The data have been analyzed in the
framework of the synthesis models for the X-ray background by Comastri
et al. (2001), and the correlation with the soft X-rays has been
investigated by Vignali et al. (2001).

The spectroscopic follow up of the HELLAS sources has been carried out
in a subsample of 118 sources out of a total of 147. 
The {\it BeppoSAX} X-ray positions have an uncertainty of about 1-1.5
arcmin depending on the distance of the sources from the axis of the
telescope. We have thus searched for optical counterparts having R
magnitude brighter than 20.5-21.0 in a circular region of 1-1.5 arcmin
of radius around the HELLAS positions. In the case of large off-axis
distances, the larger error-boxes (1.5$''$) have been used. 25 sources
have been identified with cross-correlation with existing catalogues
available from NED, and 49 have been investigated at the telescope.

We have divided the identified sources in AGN1, AGN2, galaxies with
narrow emission lines showing moderate-to-high degree of excitation
(ELG), and stars. AGN2 includes AGN1.5, AGN1.8 and AGN1.9. No
distinction has been done among radio-loud and radio quite
objects. The R-band magnitudes of these identifications have been
pushed down to the limit of obtaining less than 10$\%$ probability of
having a spurious identification due to chance coincidences for each
of the HELLAS sources (i.e. 90$\%$ reliability for the whole sample).

For AGN1 we have chosen a limit of R=21 where the surface density is
$\sim$60-70/deg$^2$.  We have chosen a limit of R=19 for AGN2.  We
have collected all the objects which are mainly starburst galaxies and
low ionization narrow emission line galaxies (LINERS) within the class
of emission line galaxies (ELG).  For these objects a limit of R=17.5
has been chosen.  In total, 63$\%$ (74/118) of our HELLAS subsample
has been searched for spectroscopic identification. 61 have been
identified: 37 AGN1, 9 AGN2, 5 ELG, 6 Clusters, 2 BL Lac, 1 Radio
Galaxy and 1 Star.

\subsection{The evolution of AGN1}

For the reasons described before we have decided, at moment, to limit
the analysis of the evolution of AGN to the AGN1 only.  
We have
assumed H$_0$=50 km/s/Mpc and either the
($\Omega_{\mathrm{m}}$,$\Omega_\Lambda$)=(1.0,0.0) or the
($\Omega_{\mathrm{m}}$,$\Omega_\Lambda$)=(0.3,0.7) cosmologies. In order to join
our data with other existing samples we have computed the 2-10 keV
luminosity for type 1 AGN assuming a typical spectral energy
distribution as computed by Pompilio, La Franca \& Matt (2000), which
is roughly approximated by a single power slope in energy with index
0.6 (${dF\over dE}\propto E^{-0.6}$) in the range 2-50 keV (i.e. up to
$z$=4).

The statistical significance of our analysis has been increased by
joining our data to other hard X-ray samples of AGN1 identified by
Grossan (1992), Boyle et al. (1998), and Akiyama et al. (2000).

We carried out a least-squares method to derive best-fit parameters
for the 2-10 keV luminosity function (LF) and its cosmological
evolution. All the fits have been tested with the bidimensional
Kolmogorov-Smirnov test (2DKS). The summary of the results is
presented in Table 1.  Two different functional forms have been
used. Following Boyle et al. (1998) and Ceballos and Barcons (1996),
we used the pure luminosity evolution (PLE) for the QSO LF. We also
used the luminosity dependent density evolution (LDDE) model similar
to the one fitted in the soft X-rays by Miyaji et al. (2000).

The local ($z$=0) QSO LF used for the PLE model has been represented by a
two-power-law:

\begin{eqnarray}
	\begin{array}{ll}

{d\Phi(L_{X}) \over dL_{X}} = A (L_X^{*(\gamma_1-\gamma_2)}) L_{X}^{-\gamma _1} & : L_X \leq L_X^*(z=0), \nonumber \\
{d\Phi(L_{X}) \over dL_{X}} = A L_{X}^{-\gamma _2} & : L_X > L_X^*(z=0),\nonumber 
	\end{array}
\end{eqnarray}

\noindent
where $L_{X}$
is expressed in units of $10^{44}{\rm erg\,s^{-1}}$.
A standard power-law luminosity evolution model was
used to parameterize the cosmological evolution of this LF:
$L_X^*(z) = L_X^*(0)(1+z)^{k}$. 

In the case of the LDDE model, as an analytical expression of the
present day ($z=0$) luminosity function, we used the smoothly
connected two power-law form:

$$ {d\Phi(L_{\rm x},z=0) \over d{\rm Log}L_{\rm x}} = A[(L_{\rm x}/L_\ast)^{\gamma_1} + (L_{\rm x}/L_\ast)^{\gamma_2}]^{-1}.$$

\noindent
The description of the evolution is given by a factor $e(L_{\rm x},z)$
such that:

$$ {d\Phi(L_{\rm x},z) \over d{\rm Log}L_{\rm x}} = {d\Phi(L_{\rm x},z=0) \over d{\rm Log}L_{\rm x}} \times e(L_{\rm x},z),$$

where
\begin{eqnarray}
  \lefteqn{e(L_{\rm x}, z) = } \nonumber \\
      & \left\{ 
	\begin{array}{ll}
	  (1+z)^{\max(0,{p1}-{\alpha}\,{\rm Log}\; 
	  {[L_{\rm a}/L_{\rm x}]})} 
	     & (z \leq z_{\rm c}; L_{\rm x}<L_{\rm a}) \\ 
	  (1+z)^{p1}
	     & (z \leq {z_{\rm c}}; L_{\rm x}\ge L_{\rm a})\nonumber \\ 
        e(L_{\rm x},{z_{\rm c}})
         \left[(1+z)/(1+{z_{\rm c}}) \right]^{p2} 
	     & (z>{z_{\rm c}}). \\
	\end{array}
       \right.
\end{eqnarray}

\begin{figure}[t]
\plottwo{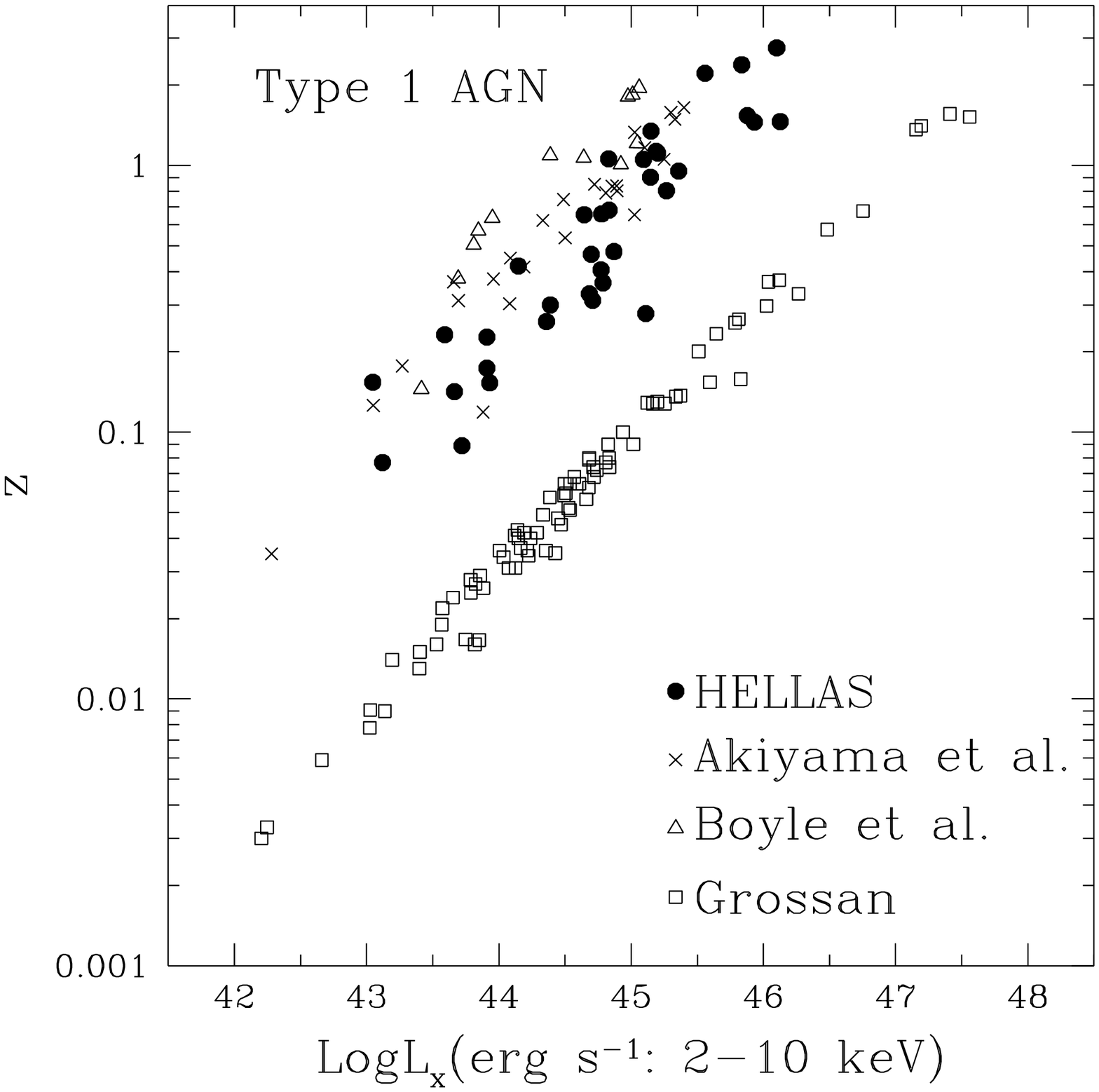}{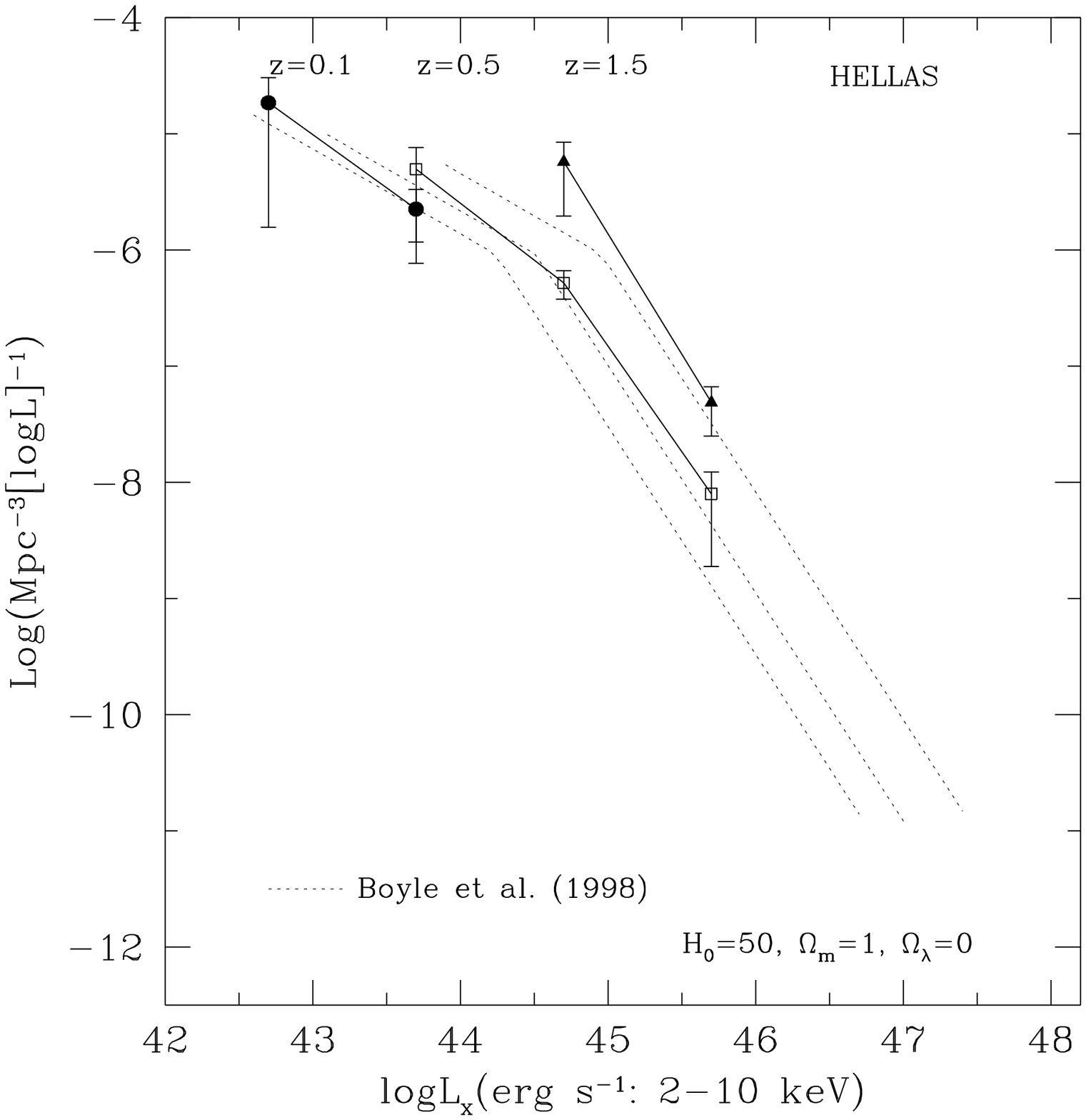}
\caption{
a) The luminosity/redshift distribution for all the 156 AGN1 used to
compute the hard X-ray 2-10 keV luminosity function.  They include 37
AGN1 from HELLAS, 82 AGN1 from Grossan, 12 AGN1 from Boyle et
al. (1998) and 25 AGN1 from Akiyama et al. (2000).
b) The luminosity function
of the 37 type 1 AGN from HELLAS.  The dotted lines show, as a
reference, the previously estimated luminosity function in the 2-10
keV from Boyle et al. (1998).  The densities have been corrected for
evolution within the redshift bins.}
\end{figure}

\begin{figure}[t]
\plottwo{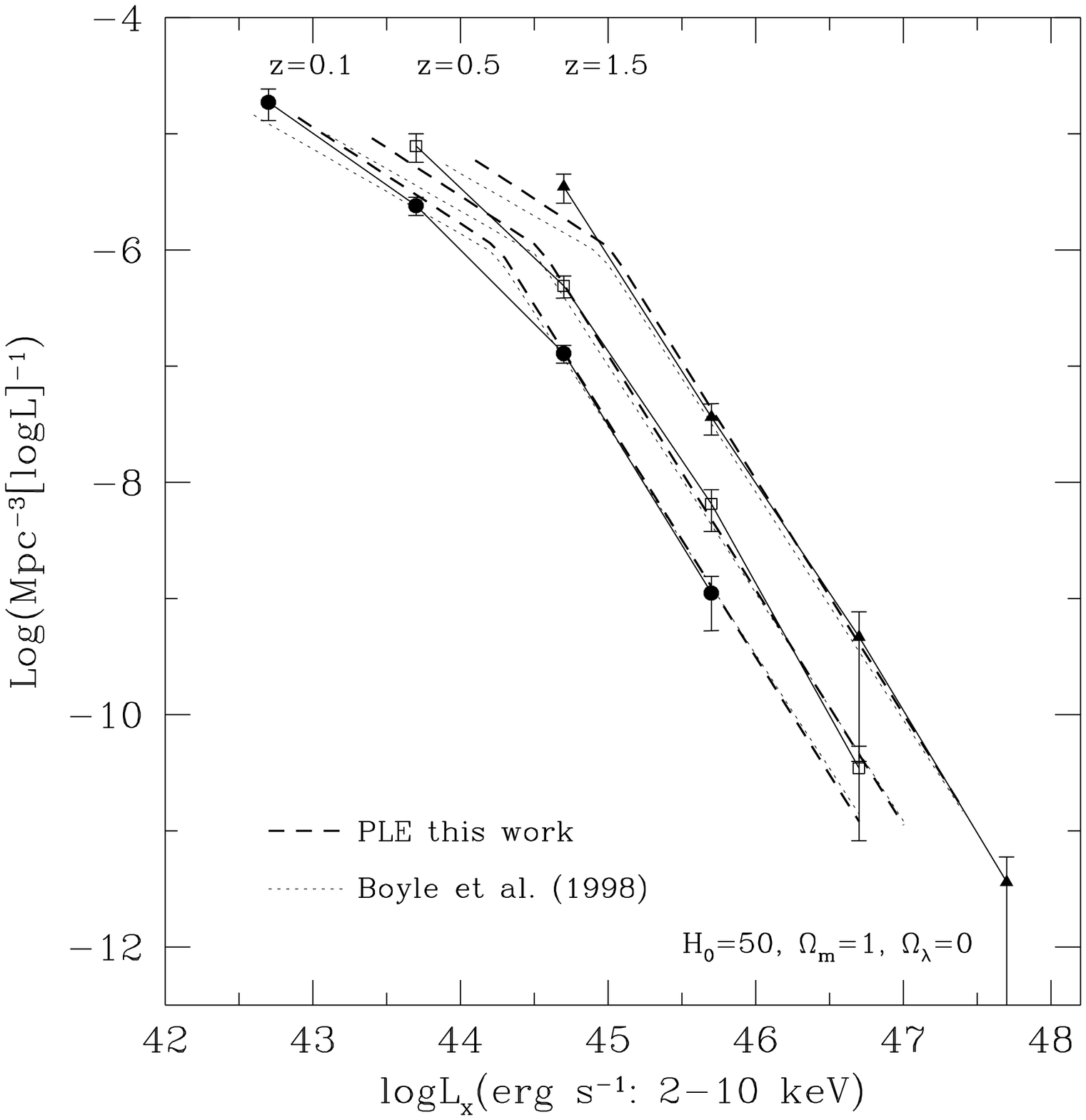}{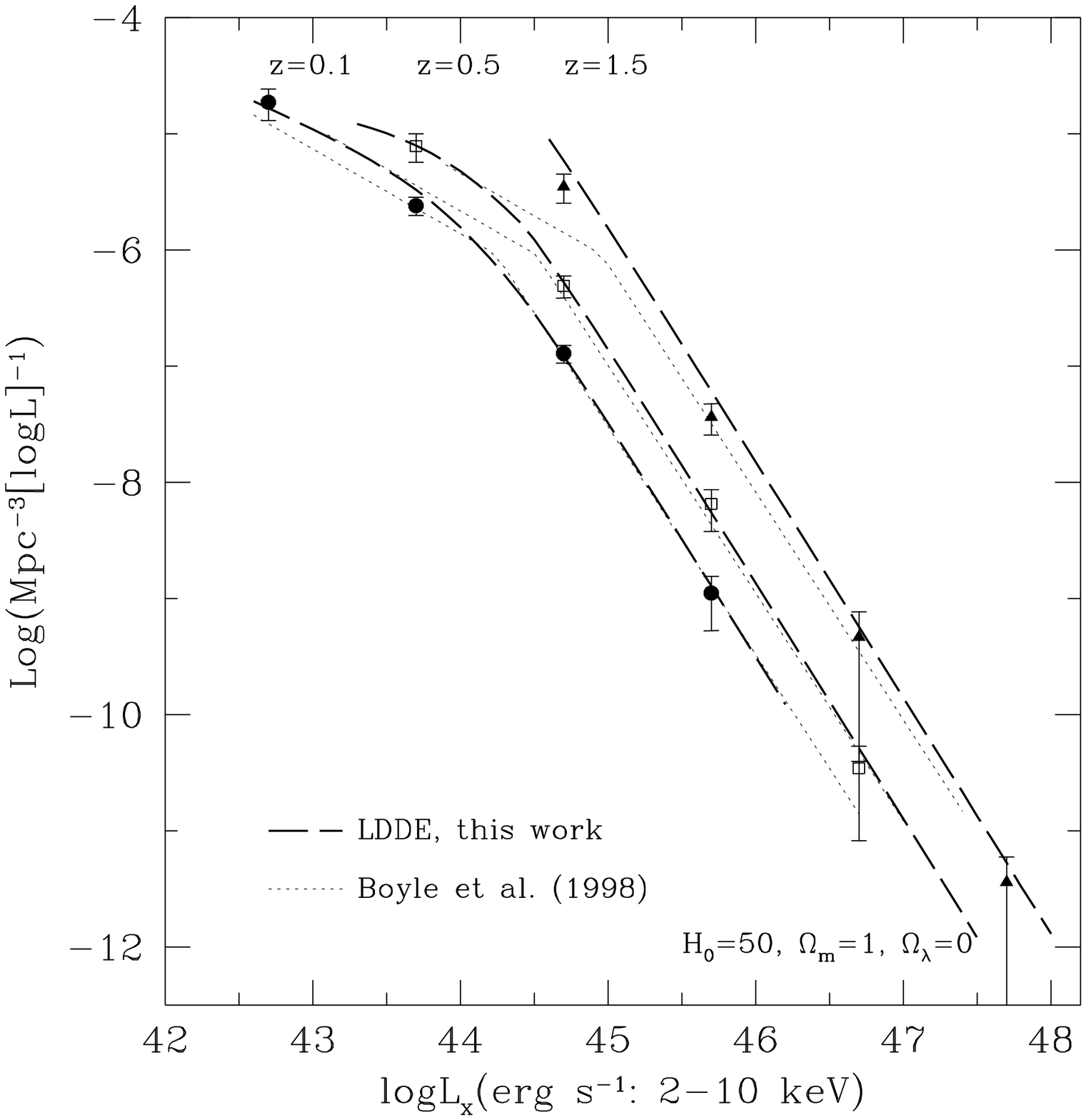}
\caption{
a) The luminosity function of the total sample of 156 AGN1 including
the 37 AGN1 from HELLAS, 82 AGN1 from Grossan, 12 AGN1 from Boyle et
al. (1998) and 25 AGN1 from Akiyama et al. (2000). The excess of faint
AGN1 at high redshifts is even more evident than in the previous
figure. The dotted line is the PLE fit (model 0, Table 1) by Boyle et
al. (1998). The dashed line is our best PLE fit (model 1, Table 1).
b) The luminosity function of AGN1 fitted with the LDDE model (model
4 in Table 1, see text).  }
\end{figure}

\noindent
We started our computation of the luminosity function in an
($\Omega_{\mathrm{m}}$,$\Omega_\Lambda$)=(1.0,0.0) 
cosmology. The only previous
estimate of the evolution of AGN1 in the 2-10 keV range was from Boyle
et al. (1998), who joined the local sample of 84 AGN1 observed by {\it
HEAO1} from Grossan (1992) with a fainter sample of 12 AGN1 observed
by {\it ASCA} (see their distribution in the
L$_X$-$z$ plane in Figure 1a).

\begin{table}[b]
\footnotesize
\caption{The 2-10 keV AGN LF parameters}
\begin{tabular}{llllllllcc}
\hline
\\
Model & $\gamma_1$ & $\gamma_2$ & Log$L^*$ & $k~or~p1$ & $z_{cut}$ & $A^a$ & $P_{2DKS}$ \cr
\hline
0) PLE from Boyle$^\ast$     & 1.73     & 2.96     & 44.16     & 2.00     & ...     & $8.2\times 10^{-7}$      & $0.04^b$ \cr
1) PLE$^\ast$                & 1.83     & 3.00     & 44.17     & 2.12     & ...     & $9.8\times 10^{-7}$      & $0.22$ \cr
2) PLE with $z_{cut}$$^\ast$ & 1.87     & 3.03     & 44.17     & 2.52     & 1.39    & $8.9\times 10^{-7}$      & $0.31$ \\
3) LDDE/a $^\ast$            & 0.62$^c$ & 2.25$^c$ & 44.13     & 5.40$^c$ & 1.55$^c$& $1.4\times 10^{-6}$ $^{c,d}$ & $0.11$ \\
4) LDDE/b$^\ast$             & 0.68     & 2.05     & 44.03     & 4.66     & 1.55$^c$& $2.0\times 10^{-6}$ $^d$ & $0.47$ \\
5) PLE$^\diamond$                & 1.93     & 2.96     & 44.23     & 2.19     & ...     & $8.6\times 10^{-7}$      & $0.48$ \\
6) PLE with $z_{cut}$$^\diamond$ & 1.92     & 2.96     & 44.22     & 2.23     & 2.31    & $8.5\times 10^{-7}$      & $0.73$ \\
\\
68$\%$ confidence errors & $^{+0.12}_{-0.16}$ &$^{+0.11}_{-0.09}$ &$^{+0.13}_{-0.15}$ & $^{+0.13}_{-0.14}$ &$^{+0.56}_{-0.25}$ & 8$\%$~~~~ \\
\hline
\hline
\multicolumn{8}{l}{($\ast$) ($\Omega_{\mathrm{m}},\Omega_\Lambda$)=(1.0,0.0); ($\diamond$) ($\Omega_{\mathrm{m}},\Omega_\Lambda$)=(0.3,0.7)}
\\
\multicolumn{8}{l}{(a) Mpc$^{-3}$($10^{44}$erg s$^{-1})^{-1}$; (b) $\chi^2$ probability; (c) fixed; (d) Mpc$^{-3}($erg s$^{-1})^{-1}$}
\end{tabular}
\end{table}

In Figure 1b the luminosity function from only our 37 AGN1 from HELLAS
in three redshift intervals ($0.0<z<0.2$, $0.2<z<1.0$, and
$1.0<z<3.0$) is shown. For comparison the best fit luminosity function
from Boyle et al. (1998) is also shown. The data have been represented
by correcting for evolution within the redshift bins as explained in
La Franca and Cristiani (1997). The data are in rough agreement with
the previous estimate of Boyle et al. (1998), but show some evidence
of an excess of faint AGN1 at redshifts larger than $z\sim$0.3. This
feature is strengthened by joining our data with the other AGN1
samples from Grossan (1992), Boyle et al. (1998), and Akiyama et
al. (2000) (see Figure 2a). These samples all together collect 156
AGN1. These data have a ($\chi^2$) 0.04 probability to be drawn from a
PLE model such as that computed by Boyle et al. (1998, see model 0 in
Table 1).  Our best fit to the whole data with a PLE model found
similar parameters as those of Boyle et al. (1998) but a slightly
larger evolution ($k=2.13$ instead of $k=2.04$) and a significantly
larger normalization (model 1). This difference is originated by the
necessity of better fitting the observed higher density of faint AGN1
at high redshifts. The 2DKS test gives a probability of 0.22 for this
fit (see Figure 2a and Table 1). A even better probability of 0.31 is
obtained if a stop in the evolution is applied at redshift
$z_{cut}=1.39$ and a larger evolution ($k=2.52$) is used (model 2).

Although our fits are already statistically adequate, our PLE model is
not able to fully describe the observed over-density of faint AGN1 at
high redshifts. As this feature is quite similar with what observed in
the soft X-rays (Miyaji et al. 2000) we tried to get a
even better fit by using the luminosity dependent density evolution
(LDDE) model similar to the one fitted in the soft X-rays.  We first
kept fixed all the parameters to the values found in the soft X-rays,
just leaving $L_{\ast}$ free to vary (model 3). The 2DKS test gives a
probability of 0.11. The value Log$L_\ast$=44.13 found by our fit
nearly corresponds to the value Log$L_\ast$=43.78 found in the soft
X-rays if a linear relation $L_X$(0.5-2 keV)=0.471$\times$L$_X$(2-10
keV) is assumed, which corresponds to the slope 0.6 of the AGN1
spectra which we used in this analysis.  Later we kept fixed just
$z_c=1.55$ and $\alpha=2.5$ to the value found by Miyaji et al. (2000)
for AGN1 only, and left all the remaining parameters free to vary
(model 4, see Figure 2b). In this case the 2DKS test gives a probability of 0.47.

\begin{figure}
\plottwo{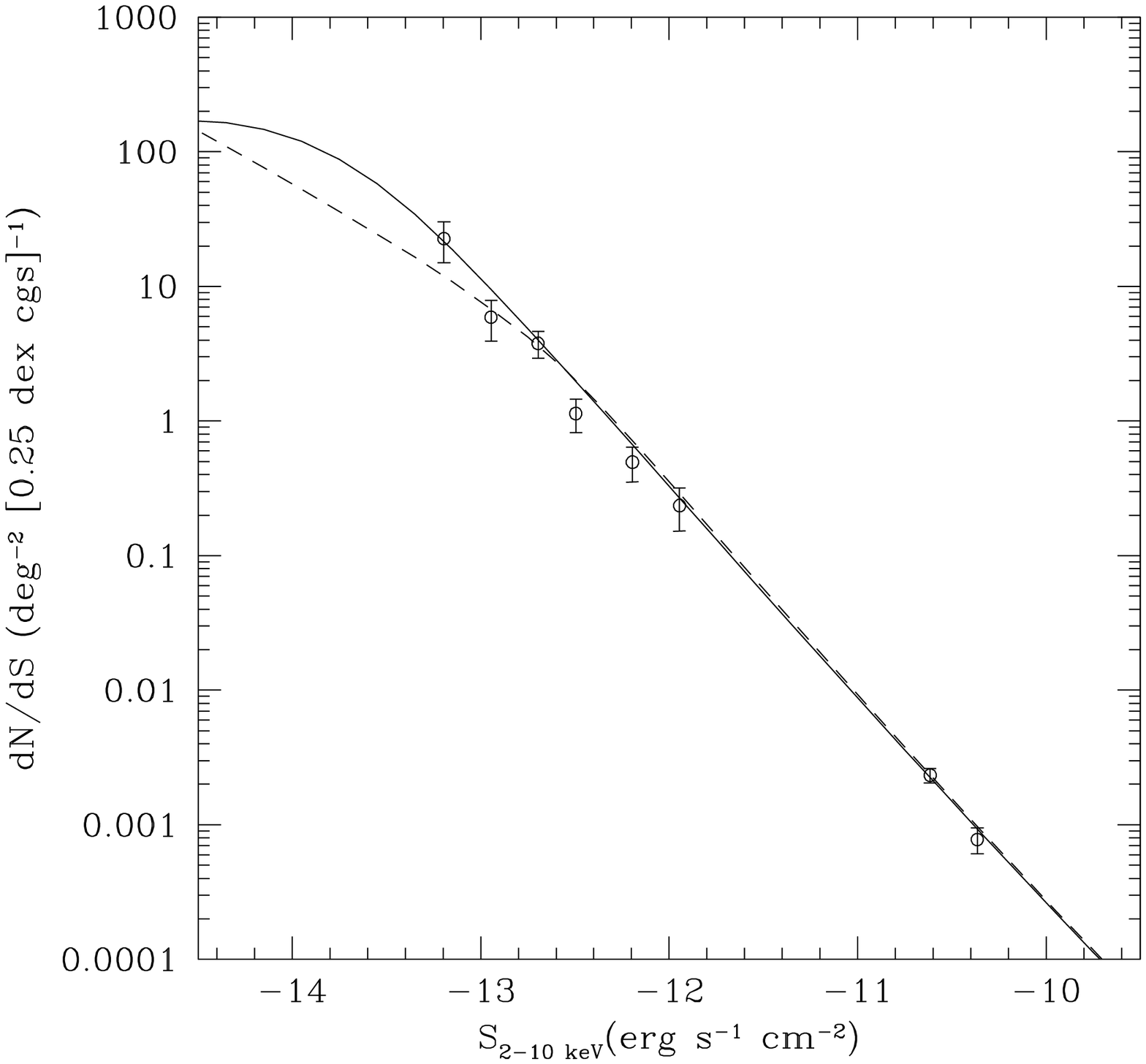}{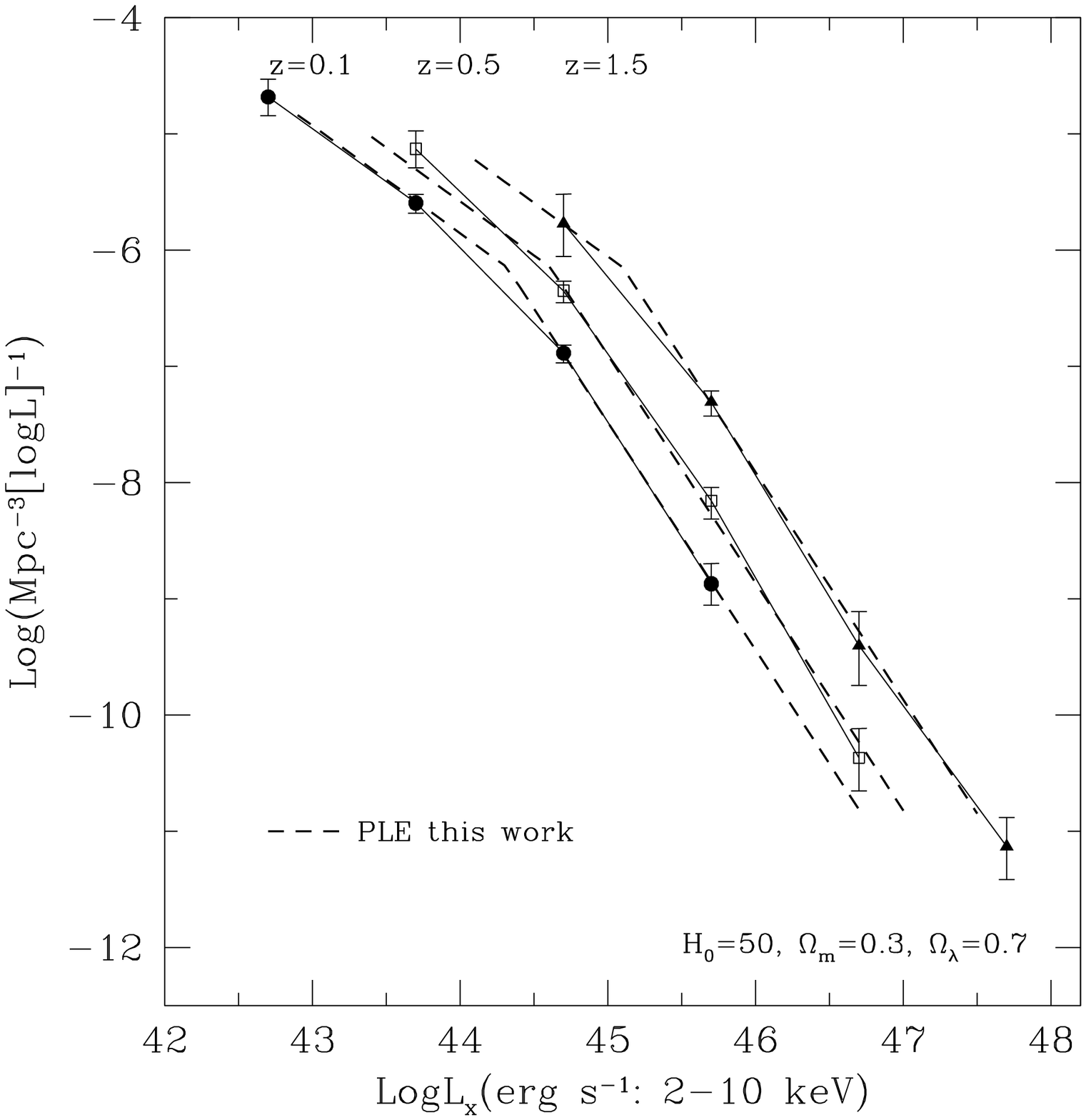}
\caption{ a)
The differential counts of AGN1. The continuous line is the expected counts for the LDDE/b
model. The dashed line is the expected counts for the PLE model with
$z_{cut}$. ($\Omega_{\mathrm{m}},\Omega_\Lambda$)=(1.0,0.0) is assumed. b) 
The luminosity function in the ($\Omega_{\mathrm{m}},\Omega_\Lambda$)=(0.3,0.7)
Universe fitted by our PLE model (model 5 in Table 1).}
\end{figure}

Although we find a better fit of the data with the LDDE model, the
existing sample is not able to distinguish between PLE and LDDE. In
Figure 3a the differential counts of the 156 AGN1 used in the
evaluation of the evolution of the LF is shown. The continuous line
is the prediction of the LDDE/b model while the dashed line is the
prediction of the PLE model with $z_{cut}$.
($\Omega_{\mathrm{m}},\Omega_\Lambda$)=(1.0,0.0) is assumed.  The two models
differentiates at fluxes fainter than $f_{2-10 keV}$$\sim$$10^{-13}
erg~cm^{-2}s^{-1}$, where the LDDE/b model predicts a higher density
of AGN1. The new upcoming fainter surveys from {\it Chandra} and {\it
XMM-Newton} will easily test which model is correct.
 
Our best fit LF LDDE/b model predicts a contribution of AGN1 to the
2-10 keV X-ray background of $I_{2-10}= 1.04\times 10^{-11}$ erg
cm$^{-2}$ s$^{-1}$ deg$^{-2}$, while the PLE model with $z_{cut}$
predicts $I_{2-10}= 0.76\times 10^{-11}$ erg cm$^{-2}$ s$^{-1}$
deg$^{-2}$. These values correspond to a fraction of 53$\%$ and 39$\%$
respectively of the 2-10 keV X-ray background (we used $I_{2-10}=
1.95\times 10^{-11}$ erg cm$^{-2}$ s$^{-1}$ deg$^{-2}$ from Chen,
Fabian and Gendreau 1997).

If an ($\Omega_{\mathrm{m}}$,$\Omega_\Lambda$)=(0.3,0.7) cosmology is assumed the
PLE models obtain an ever better representation of the data in
comparison with what found in the
($\Omega_{\mathrm{m}}$,$\Omega_\Lambda$)=(1.0,0.0) Universe (see Table 1, models
5 and 6).  The 2DKS probability is 0.73 and 0.48, with and without the
introduction of the $z_{cut}$ parameter respectively. In this case
even the simple PLE model obtains a quite good fit of the data (Figure
3b), and the introduction of the $z_{cut}$ parameter is necessary to
stop the evolution only at redshifts larger than 2.3. However our data
contain not enough AGN1 at redshift larger than 2 in order to obtain
an accurate measure of the $z_{cut}$ parameter (see Figure 1a), and
the errors in this parameter are quite large.

\section{ELAIS: The evolution of AGN1 in the MIR (15$\mu$m)}

\subsection{The ELAIS survey}

The European Large Area ISO Survey (ELAIS) is the largest single open
time project conduced by ISO, mapping an area of 12 deg$^2$ at 15$\mu
$m with ISOCAM and at 90$\mu $m with ISOPHOT.  Four main fields were
chosen (N1, N2, N3 in the north hemisphere and S1 in the south) at
high Ecliptic latitudes ($|\beta|>40^{\scriptscriptstyle{o}}$).

An initial catalog for S1 (J2000,$\, \alpha:
00^{\scriptscriptstyle{h}}34^
{\scriptscriptstyle{m}}44^{\scriptscriptstyle{s}}\, ,\delta: {-43}^
{\scriptscriptstyle{O}}34^{\scriptscriptstyle{'}}44''$), covering and
area of 3.96 deg$^2$, was produced using the Imperial College data
reduction technique (``Preliminary Analysis'', Serjeant et al., 2000).
Optical identifications were possible thanks to an extensive R-band
CCD survey, performed with the ESO/Danish 1.5m telescope.  The
spectroscopic follow-up program, carried out at the AAT at the AAO and
the 3.6m/NTT at \mbox{ESO/La Silla}, of 114 sources in S1 fainter than
R$\sim$17.0 provided the sample presented here.

The ELAIS-S1 field has also been completely covered in the radio at
1.4 GHz down to 0.3 mJy (Gruppioni et al., 1999), and 50\% covered in
the X-rays with BeppoSAX (Alexander et al., 2001).  Now the Final
Analysis has been completed by the ELAIS team in Bologna. The
resulting complete catalog includes more that 450 sources, with fluxes
at 15$\mu$m down to 0.5 mJy and selected with S/N$>$5 (Lari et al.,
2001).

For many years there has been a gap in flux between the brighter
samples ($>$300 mJy), coming from the IRAS surveys and covering large
sky areas, and the deep/pencil-bin surveys carried out by ISO at much
fainter fluxes ($<$1 mJy).  In the integral counts derived from the
final analysis (Figure 4a, Gruppioni et al., in prep) we see how the
ELAIS survey fills the whole flux range between these two regimes.

\begin{figure}
\plottwo{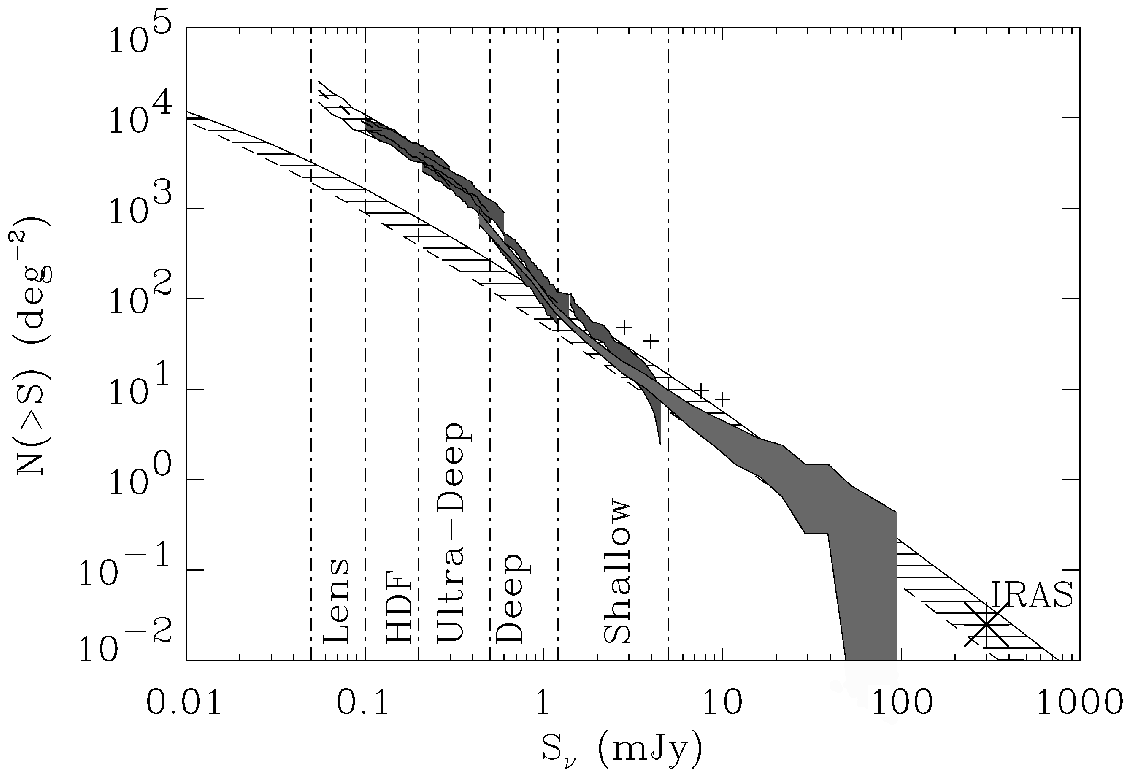}{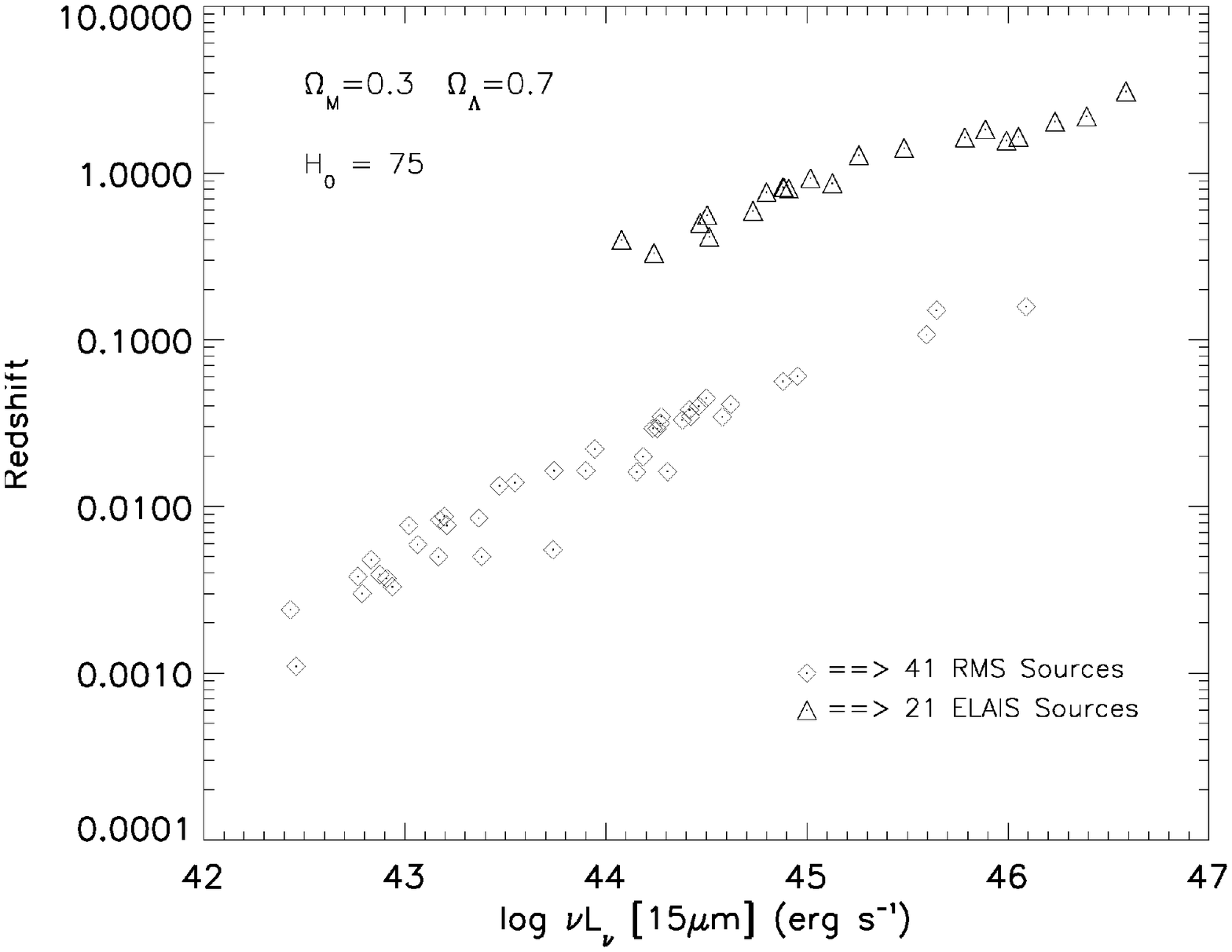}
\caption{a) Integral Counts at 15$\mu$m from the Final Analysis of the 
ELAIS-S1 Region (light grey shaded area). Also plotted the faintest
IRAS surveys and the deep ISOCAM surveys (dark grey shaded areas).  b)
The luminosity/redshift distribution of our 21 ELAIS-S1 and 41 RMS
AGN1s that entered into the computation.}
\end{figure}

\noindent
The nature of the objects identified by the spectroscopic follow-up
is:

{\bf --} High star-forming galaxies (45$\%$);

{\bf --} AGNs (type 1 $\&$ 2) represent 30$\%$;

{\bf --} 15$\%$ of galaxies dominated by absorption lines;

{\bf --} 4$\%$ of late-type stars;

{\bf --} 6$\%$ of unclear classification (AGN2, Starburst, LINER).

\begin{center}
\footnotesize
\begin{tabular}{cccccc}
\multicolumn{6}{l}{Table 2.~~~ The percentages of all objects with emission lines}
\\
\hline
\\
 & &  AGN1  &  AGN2  &  ELG  &  LINERS  \\
\hline
\textbf{ELAIS-S1} & $\%$    & 20  & 10  & 50  &2 \\
 15$\mu$m         & $<$z$>$ & 1.4 & 0.3 & 0.2 &0.3 \\
\\
IRAS     & $\%$  &6&17&54&23 \\
 12$\mu$m        & $<$z$>$     &0.04&0.015&0.014&0.007 \\
\hline
\hline
\multicolumn{5}{l}{Note - IRAS subsample from Alexander \& Aussel (2000)}
\end{tabular}
\end{center}

The observed ratio of the number of type 2 with respect to type 1+2 AGNs
is 1/3. But this result is mainly an artifact due to the
different selection functions of the sample for AGN1 and AGN2. The
optical selection introduced in the spectroscopic follow-up ($\mathrm
17.0$$<$$R$$<$$20.0$) seems to be at the origin of the lack of type 2 AGNs
beyond $z\sim$0.5. As a matter of fact, in Figure 5a a trend is
observed for AGN2 with optical magnitudes being fainter at larger
redshifts. In Figure 5b the redshift distributions of AGN1 and AGN2
are shown. In the redshift bin where type 2 AGNs are observed
($z$=0.0-0.5), the ratio AGN2/AGN1 becomes $\sim$4, similar to the
predictions of the standard unification model.  The follow-up campaign
to be carried out at September/November 2001 with the ESO telescopes
will verify if this trend is seen at fainter and brighter optical
fluxes.

\begin{figure}
\plottwo{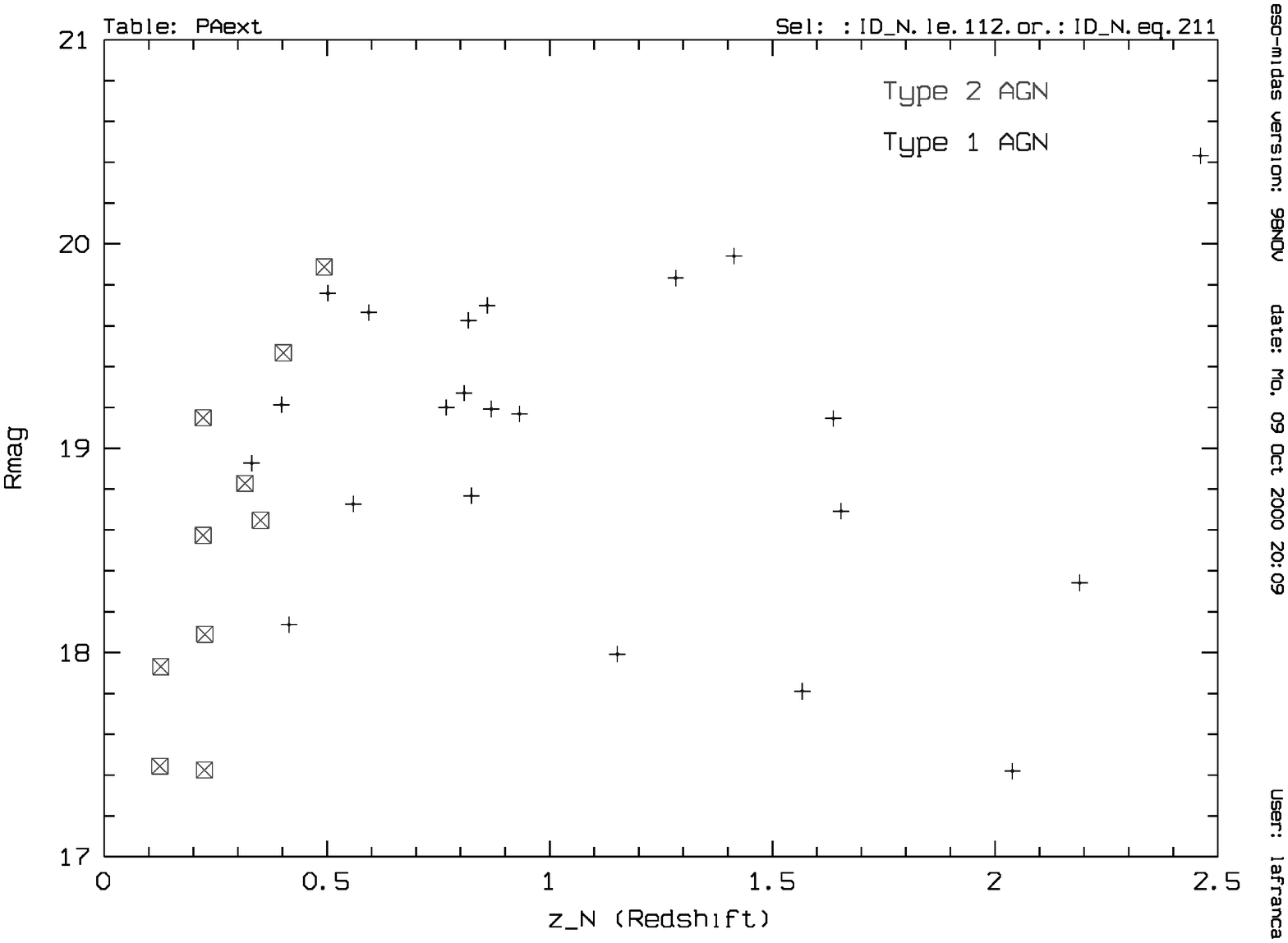}{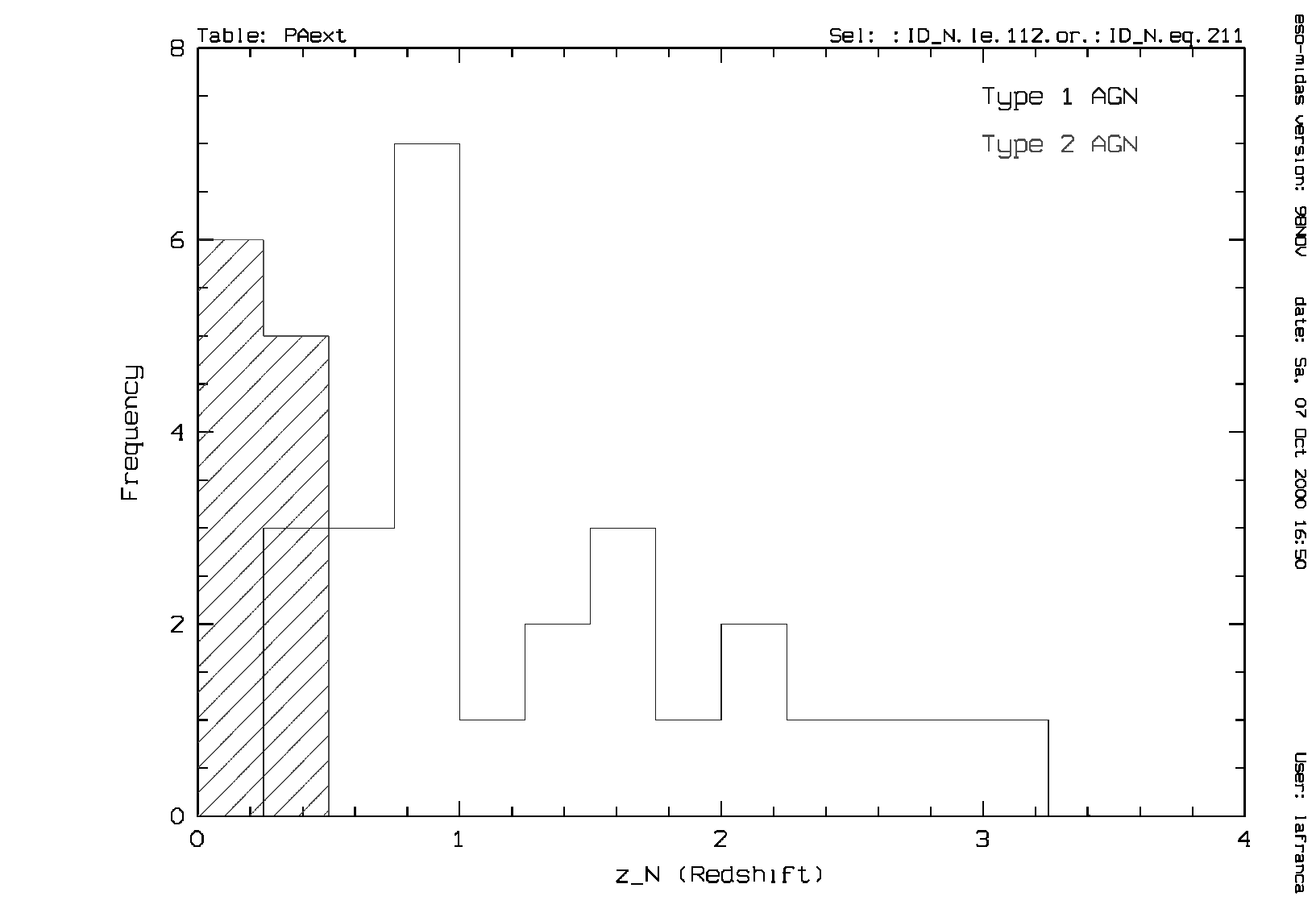}
\caption{a) Distribution of AGN1 (plus symbols) and AGN2 (crossed squares) in the
redshift-Rmag space.  While the R-magnitude distribution is constant
with a large spread for AGN1, AGN2 show a trend with redshift,
becoming fainter than R=20.0 at $z\sim$0.5. b) Distribution in the
redshift space of AGN1 and AGN2 (shaded area). If we consider the
redshift bin $z$=[0.0,0.5] the ratio of AGN2/AGN1 becomes 11/3$\sim$4.
}
\end{figure}

\subsection{The Evolution of AGN1}

We have assumed H$_0$=75 km/s/Mpc and
a ($\Omega_{\mathrm{m}}$,$\Omega_\Lambda$)=(1.0,0.0) cosmology.
Our ELAIS preliminary sample of 21 AGN1 is statistically
significant enough to compute a first estimate on the evolution of these
objects in the \mbox{Mid-IR}. The mean SED from Elvis et al. (1994)
for radio quiet QSOs was adopted as a good representation of our
sources. $K$-corrections were computed for each of the two different
filters: ISOCAM-LW3 at 15$\mu$m and IRAS 12$\mu$m.

A subsample of AGN1 was extracted from the catalog of Rush, Malkan
\& Spinoglio (1993), as representative in the
local universe of this type of objects. The catalog consists of a
sample of galaxies selected at 12$\mu$m from the IRAS Point Source
Catalog PSCv2, and is complete down to \mbox{0.3 Jy.}  With the
computed $K$-correction, 15$\mu$m $\nu L_{\nu}$ luminosities (L$_{15}$) 
were derived. Figure 4b represents, in the luminosity-redshift space,
all AGN1 coming from ELAIS and RMS that have been used in this
analysis.

Similarly to what found in the optical and in the X-rays we adopted a
smooth double power-law for the space density distribution of QSOs in
the local universe ($z$=0):

$$ {d\Phi(L_{\rm IR},z=0) \over d{\rm Log}L_{\rm IR}} = 
\rm C\;\Big[(L_{\rm IR}/L_\ast)^{\alpha} + (L_{\rm IR}/L_\ast)^{\beta}\Big]^{-1},$$

the standard luminosity evolution (PLE) has been adopted: $ L_{\rm
15}(z) = L_{\rm 15}(0)(1+z)^{k}$.
A parametric, unbinned maximum likelihood method was used to fit the
evolution and luminosity function parameters simultaneously (Marshall
et al., 1983) at 15$\mu$m. Since ELAIS identifications were not only
flux limited at 15$\mu$m, but also in their R-band magnitude (17.0 $<$
R $<$ 20.0), a factor $\mathbf{\Theta}$ was introduced in the function
'S' to be minimized

\begin{center}
\begin{displaymath}
\mathrm{S}=-2 \,\sum_{i=1}^{N} ln[\Phi(z_{i},L_{i})] + \int\!\!\!\int 
\Phi(z,L) \Omega(z,L) \, \mathbf{\Theta} \, \mathrm \frac {\textstyle dV}
{\textstyle dz}dzdL,
\end{displaymath}
\end{center}
to correct for incompleteness, and only applied to the ELAIS sample.
This factor $\mathbf{\Theta}$ represents the probability that a source
with a given luminosity at 15$\mu$m (L$_{15}$) has a R-magnitude
(L$_{\rm R}$) between the limits of the sample (17.0$<$R$<$20.0),
$$\mathbf{\Theta}\mathrm (z,L) \,(17.2<R<20.0 \, \mid
\mathrm{L}_{15})\, .$$ and was derived taking into account the 1$\sigma$
internal spread in the SED.

From the total available list of AGN1 of the sample, the number 
of objects that entered the computation was defined as:
\begin{itemize}
\item[-] {RMS: sources with $\mathrm F_{12\mu m} \ge 300 \;mJy$, 41 sources}
\item[-] {ELAIS: sources with $\mathrm F_{15\mu m} \ge 1\;mJy$, and 
$\mathrm 17.0 < R < 20.0$, 21 sources}
\end{itemize}

\begin{figure}
\begin{center}
\plotfiddle{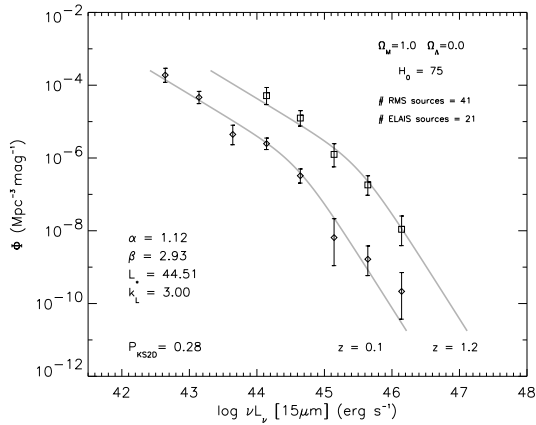}{5.6cm}{90}{35.0}{35.0}{180}{-20}
\caption{PLE fit to our 63 total sources (RMS + ELAIS-S1).
The densities have been corrected for evolution within the redshift bins.}
\end{center}
\end{figure}

The resulting estimate of the LF of AGN1 at 15$\mu$m with its
parameters is shown in Figure 6. The probability that our data have
been drawn from the fitted PLE model is 0.28, as given by the 2DKS
test.  The PLE model adopted here could not be sufficient to represent the
space density of our sources, some density evolution may be
required. Currently we are studying different parameterizations taking
into account some degree of density evolution.

For the derived PLE model the contribution of AGN1 to the CIRB at
15$\mu$m is $\nu I_{\nu}=5.2 \mathrm \times 10^{-11}\;W m^{-2}
sr^{-1}$ which corresponds \mbox{$\sim 2\%$} of the lower limit to the
CIRB calculated by Altieri et al. 1999 ($\nu I_{\nu}=3.3\times
10^{-9}$ W m$^{-2}$ sr$^{-1}$) in this band (see also Hauser \& Dwek 2001;
Hauser 2001).  At maximum, the total contribution of AGNs to the
background at 15$\mu$m can be as high as $\nu I_{\nu}=\mathrm
2.6\times10^{-9} \;W m^{-2} sr^{-1}$ ($\sim$10$\%$ of the background
measured by Altieri et al. under the extreme assumptions: 1) that AGN2
are as much bright as AGN1 at 15 $\mu$m, 2) that the ratio of type 2
to type 1 AGN is 4 at all redshifts, and 2) that AGN2s evolved with
the same LF of AGN1.

\subsection{Conclusions}

Thanks to the spectroscopic campaigns carried out on the HELLAS and
ELAIS catalogues, we have been able to build up statistical significant
samples of AGN1 and estimate the evolution of their LF in the hard
X-rays and IR.

In the hard X-rays we have been able to double the number of hard
X-ray AGN1 available for statistical analysis at fluxes in the range
$f_{2-10 keV}\sim 10^{-12}- 10^{-13.5} erg~cm^{-2}s^{-1}$. In total we
have used 74 AGN1 at these fluxes, which combined with the local
sample of Grossan (1992) have allowed to show directly the shape of
the LF of AGN1 as function of redshift. In the IR, our sample of 21
AGN1 has allowed the first estimate of the evolution of the LF at
15$\mu$m.

Both LFs are fairly well represented by a double-power-law-function
with a significant cosmological evolution according to a PLE model
with $L(z)$$\propto$$(1+z)^k$. However, there are some evidences of
density evolution (LDDE) at faint luminosities.

\begin{acknowledgements}

Based on observations collected at the European Southern Observatory,
Chile, ESO N$^{\circ}$: 62.P-0783, 63.O-0117(A), 64.O-0595(A),
65.O-0541(A).  We thank the BeppoSAX SDC, SOC and OCC teams for the
successful operation of the satellite and preliminary data reduction
and screening.  This research has made use of the NASA/IPAC
Extragalactic Database (NED) which is operated by the Jet Propulsion
Laboratory, California Institute of Technology, under contract with
the National Aeronautics and Space Administration.  This research has
been partially supported by ASI contracts ARS-99-75, ASI 00/IR/103/AS,
ASI I/R/107/00, MURST grants Cofin-98-02-32, Cofin-99-034,
Cofin-00-02-36, Chandra X-ray center grant G01-2100X, and a 1999 CNAA
grant.
\end{acknowledgements}

\end{document}